\documentclass[conference, pdftex]{IEEEtran} 

\usepackage[dvips,pdftex]{graphicx}				
\usepackage[numbers,sort&compress]{natbib} 	
\usepackage{verbatim} 
\usepackage{hyperref}
\usepackage[utf8]{inputenc}			
\usepackage{algorithm}				
\usepackage{algpseudocode}			
\usepackage{comment}	
\usepackage{amssymb,amsfonts,amsmath,amsthm}
\usepackage[position=bottom]{subfig}
\usepackage{url}
\usepackage{multirow}
\usepackage{morefloats}
\usepackage{float}
\usepackage{placeins}
\usepackage{afterpage}

\begin{document} 

\title{Network Coding as a \\ WiMAX Link Reliability Mechanism}

\author{\IEEEauthorblockN{S. Teerapittayanon, K. Fouli, M. M\'{e}dard,\\ M.-J. Montpetit, X. Shi \vspace{0.05in}}
\IEEEauthorblockA{ Research Laboratory of Electronics (RLE)\\ Massachusetts Institute of Technology (MIT)\\ Cambridge, MA 02139\\ \{steerapi, fouli, medard, mariejo, xshi\}@mit.edu}
\and
\IEEEauthorblockN{ \\ I. Seskar \vspace{0.05in}}
\IEEEauthorblockA{ WINLAB \\ Rutgers University\\Piscataway, NJ 08854\\seskar@winlab.rutgers.edu}
\and
\IEEEauthorblockN{ \\ A. Gosain \vspace{0.05in}}
\IEEEauthorblockA{ Raytheon BBN Technologies\\Cambridge, MA 02138\\agosain@bbn.com}

}

\maketitle

\begin{abstract}

We design and implement a network-coding-enabled reliability architecture for next generation wireless networks. Our network coding (NC) architecture uses a flexible thread-based design, with each encoder-decoder instance applying systematic intra-session random linear network coding as a packet erasure code at the IP layer, to ensure the fast and reliable transfer of information between wireless nodes. 

Using Global Environment for Network Innovations (GENI) WiMAX platforms, a series of point-to-point transmission experiments were conducted to compare the performance of the NC  architecture to that of the Automatic Repeated reQuest (ARQ) and Hybrid ARQ (HARQ) mechanisms. At the application layer, \textit{Iperf} and UDP-based File Transfer Protocol (\textit{UFTP}) are used to measure throughput, packet loss and file transfer delay. In our selected scenarios, the proposed architecture is able to decrease packet loss from around 11-32\% to nearly 0\%; compared to HARQ and joint HARQ/ARQ mechanisms, the NC architecture offers up to 5.9 times gain in throughput and 5.5 times reduction in end-to-end file transfer delay.  Our experiments show that network coding as a packet erasure code in the upper layers of the protocol stack has the potential to reduce the need for joint HARQ/ARQ schemes in the PHY/MAC layers, thus offering insights into cross-layer designs of efficient next generation wireless networks.

\end{abstract}
\begin{keywords}
ARQ, GENI, HARQ, Network Coding, WiMAX
\end{keywords}

\section{Introduction}

The growing market of mobile devices is placing increasing demands on wireless networks. Indeed, at the end of 2009, the number of mobile phone subscribers exceeded 4.6 billion worldwide \cite{RefWorks:26}, and the global mobile data traffic has been predicted to double every year through 2014 \cite{RefWorks:27}. As a result, a crucial challenge for next generation wireless networks is to cope with the rapid increase in multimedia traffic with minimal impact on equipment complexity \cite{RefWorks:27}.

Network Coding (NC) was originally proposed to maximize the capacity of a wired network \cite{ahlswede:nc, koetter2003algebraic}. It enables nodes to combine or separate transient bits, packets, or flows through coding and decoding operations, in addition to storing and forwarding. In a wireless setting, NC adapts to the dynamics of the network topology, an essential feature of mobile wireless networks \cite{medard2011network}. Numerous studies have shown that the use of NC in Wireless Local Area Networks (WLANs) significantly enhances network throughput, robustness, and security \cite{RefWorks:2,RefWorks:21}. Notably, 3-4x throughput gains were demonstrated experimentally in a WiFi context through the use of simple binary network codes \cite{katti2006xors}. Random Linear Network Coding (RLNC) \cite{ho:rlnc,chou2003practical}, where the NC coefficients are selected randomly over a given Galois field, has proven particularly effective in optimizing network resource consumption in WLANs \cite{RefWorks:30,RefWorks:31}. NC may be applied across the OSI model \cite{RefWorks:33} from the physical \cite{RefWorks:30} to the network and application layers \cite{RefWorks:39}. 

Despite the demonstrated effectiveness of NC in WLANs, NC for Wireless Metropolitan Area Networks (WMANs) has gained attention only recently, as the telecommunication industry moves toward next generation wireless networks such as 4G Worldwide Interoperability for Microwave Access (WiMAX) \cite{andrews2007fundamentals} and 4G Long Term Evolution (LTE)-Advanced \cite{ghosh2010fundamentals}. 4G requires stationary speeds of 1 Gbps and mobile speeds of 100 Mbps, while 3G requires stationary speeds of 2 Mbps and mobile speeds of 384 Kbps \cite{RefWorks:41}. That is, 4G requires 500 and 260 times faster speeds than 3G in stationary and mobile cases, respectively. Thus, the need for low-cost performance-multiplying technologies such as NC is expected to become significant for WMANs in the near future. 

In this work, we design and implement an NC-enabled reliability architecture in a WiMAX platform provided by the Global Environment for Network Innovations (GENI) project. 
WiMAX offers robust, reliable, and cost-effective delivery of broadband services in metropolitan and rural areas \cite{andrews2007fundamentals}. To alleviate the impact of wireless errors on network performance, WiMAX adopts two retransmission mechanisms: Automatic Repeated reQuest (ARQ) at the upper MAC layer, and Hybrid ARQ (HARQ) at the lower MAC and PHY layers. In the proposed NC architecture, instead of using either or both of these retransmission mechanisms, we apply systematic intra-session random linear network coding as a packet erasure code at the IP layer. In particular, we consider a flexible thread-based design, where parallel encoding-decoding instances are put in place to ensure reliability is achieved without incurring significant delay.

The choice of the NC implementation layer is crucial. While NC is applicable across the OSI model, the choice of the convergence sublayer stands out for a number of reasons. First, additional performance gains at the physical layer are onerous, since existing coding schemes have achieved near-optimal efficiency levels. In contrast, NC may yield important gains when integrated within the transport and MAC sub-layers, as demonstrated by extensive recent studies \cite{Kim_CTCP,RefWorks:39, Kulkarni_ICN, Fouli_NC_in_NGPON, katti2006xors}. In the context of WMANs, transport and MAC functions are performed at the convergence and MAC sub-layers. Now, the current context for higher Internet layers (i.e., TCP/IP) is extremely dynamic. This is essentially due to the sensitivity of TCP's congestion control to the variety of transmission environments (e.g., wireless, satellite, optical long-haul, etc.), leading to the emergence of a number of alternative competing transport protocols \cite{SPDY_Protocol, CUBIC} and enhancements \cite{TCP_Survey}. This trend is compounded by the emergence of IPv6. NC may therefore benefit from the continuity offered by industrial standards such as WiMAX and LTE: In the context of WMANs, the application of NC at the convergence sub-layer would serve all supported traffic and would be independent from likely technology and protocol shifts at higher layers. In this work, we resort to an IP-based implementation since the convergence sublayer is not accessible in the GENI platform.

A series of point-to-point transmission experiments are conducted to compare the performance of our architecture to that of the HARQ and ARQ mechanisms. Since the GENI WiMAX base stations (BSs) only support chase combining (CC) HARQ, only CC-HARQ is considered in our study. At the application layer, \textit{Iperf} and UDP-based File Transfer Protocol (\textit{UFTP}) are used to measure throughput, the percentage of packet loss, and file transfer delay. In our selected scenarios, the proposed architecture substantially decreases packet loss from around 11-32\% to nearly 0\%. Compared to the HARQ and joint HARQ/ARQ mechanisms, the NC architecture offers up to 5.9 times gain in throughput and 5.5 times reduction in end-to-end file transfer delay. Our experimental setups were limited by our ability to access and configure the GENI WiMAX platform. Nonetheless, our initial assessment of the NC reliability architecture illustrates its potential advantages over the HARQ/ARQ scheme, and offers exciting opportunities for further investigation. 

One way of interpreting the potential advantages of the proposed NC architecture over HARQ/ARQ is to view the latter as an a posteriori repetition code adaptation mechanism, with rates determined by the number of reactive retransmissions for each unit of data. Since retransmissions are packet specific, the rate granularity is low, and the maximum rate is small. By comparison, NC formulates unique packets into equivalent degrees of freedom, offering three advantages as an code adaptation scheme. First, coded packets can be sent a priori, in expectation of packet losses, thus reducing the effect of large round trip times in ARQ. Second, each newly received degree of freedom can make up for any previously lost packet, thus leading to rate adaptation in steps of 1/block-size, where a block is the group of data packets coded together. Three, HARQ/ARQ relies heavily on the acknowledgment process, thus is prone to ACK/NACK errors, delays, and losses, which in turn can result in inefficient retransmission of correctly received packets. NC is less sensitive, since each transmitted coded packet is a new degree of freedom that can be useful in decoding. The combination of proactive transmissions, rate adaptation with a finer granularity, and robustness to ACK losses makes NC an efficient alternative reliability mechanism. It is also more in-line with the ever increasing speed and performance of a priori adaptative modulation and coding at the PHY layer.

The remainder of this article is organized as follows. Section \ref{sec:related_work} is an overview of NC-based HARQ/ARQ alternatives and enhancements. Section \ref{sec:architecture} describes the NC-based reliability architecture. Section \ref{sec:setup_and_metrics} considers the experimental setup and introduces the performance metrics. Section \ref{sec:results} illustrates and discusses the main results. Finally, Section \ref{sec:conclusions} concludes the paper.

\section{Related Work \label{sec:related_work}}

WiMAX is one of the two major WMAN standards \cite{ieee2005ieee}, along with LTE. It adopts two retransmission mechanisms for reliability: ARQ at the upper MAC layer and HARQ at the lower MAC and PHY layers. In ARQ, block retransmissions are processed independently; in HAQR, Forward Error Correction (FEC) and ARQ are combined, and subsequent retransmissions of a given information block are jointly processed with the original block. The two extensively investigated implementations of HARQ are Chase Combining (CC) and Incremental Redundancy (IR) \cite{el2004performance}. In WiMAX, both the HARQ and ARQ feature can be turned on, leading to a joing HARQ/ARQ setup. Observe that retransmissions under HARQ and/or ARQ require the reception of positive (ACK) or negative (NACK) acknowledgment messages for each block, hence compounding overhead. 

Network Coding (NC) is initially shown to be a capacity-achieving coding scheme for multicast in wired networks \cite{ahlswede:nc, koetter2003algebraic}. A number of analytical studies then propose NC as a stand-alone information-theoretic technique to improve throughput and retransmission efficiency in wireless networks. Lun et al. \cite{lun2008coding, lun2005further} show that RLNC is capacity-achieving for multicast connections in packet erasure networks. Dana et al. \cite{dana2006capacity} derive the capacity for a class of wireless erasure networks with no interference at reception and show that linear coding suffices to achieve the capacity region. Ghaderi et al. \cite{ghaderireliability} analytically quantify the reliability gain of NC for wireless multicast and show that NC asymptotically achieves performances similar to that of rateless erasure coding. Sundararajan et al. \cite{kumar2008arq} theoretically extend ARQ with RLNC \cite{ho:rlnc,chou2003practical}, while Nguyen et al. \cite{nguyen2009wireless} provide results comparing the bandwidth efficiency of RLNC to that of ARQ. In addition, Pu et al. \cite{pu2008performance} develop an information-theoretic performance bound to predict the coding gains of CC-HARQ in broadcast settings. 

Algorithmically, the combination of packet retransmissions in single-hop multiple unicast and multicast settings is among the earliest and most widely studied schemes using NC as a wireless reliability mechanism. We call such a scheme \textit{retransmission coding}. Jolfael et al. \cite{jolfaei1993new} apply XOR retransmission coding to ARQ in a point-to-multipoint connection over broadcast links, while Yong et al. \cite{yong2000xor} consider the multicast setting. Larsson et al. \cite{larsson2006multi,larsson2007analysis} study XOR retransmission coding for multi-user ARQ in multiple unicast settings and suggest that linear coding in larger fields may be used. Larsson et al. \cite{larsson2008multicast} also consider adaptive linear NC for ARQ in multicast settings, where coding coefficients are adaptively selected from a sufficiently large finite field. Note that earlier schemes closely related to NC exist. Metzner \cite{metzner1984improved}, for instance, presented a packet coding retransmission scheme for single-hop broadcast  whereby the retransmitted frame is built through XORing the NACKed frames of various receivers. 

Recently, practical NC-based retransmission algorithms have also been proposed, often using simulations to measure performances. MAC RLNC (MRNC) \cite{jin2008random, jin2008adaptive} uses RLNC at the MAC layer, where data blocks are segmented and coded together. N-in-1 NC \cite{li2010n} extends MRNC by coding over more than one block for retransmissions. The authors in \cite{li2010n} report a throughput gain of up to 106\% over conventional CC-HARQ. In addition, Manssour et al. \cite{manssour2011unicast} propose a retransmission scheme for wireless unicast using a combination of channel coding and NC, showing 68.75\% throughput gains compared to CC-HARQ. Qureshi et al. \cite{qureshi2010} present BENEFIT, an efficient retransmission algorithm for single-hop wireless multicast networks based on NC with conditional retransmission and reduced decoding times. While all previous contributions apply network coding digitally, SYNC \cite{yun2011towards} considers symbol-level network coding at the physical layer, thus making use of corrupted packets.

Some studies also combine retransmission coding with HARQ (NC-HARQ) in single-hop wireless networks \cite{tran2008joint}. NC-HARQ uses XOR retransmission coding in conjunction with FEC, thus, in effect, combining network and channel coding. Thobaben et al. \cite{thobaben2008joint} and Larsson et al. \cite{larsson2010analysis} consider NC-HARQ for multi-user HARQ in multiple unicast settings. Peng et al. \cite{peng2010research} consider NC-HARQ in both broadcast and unicast scenarios. Tran et al. \cite{tran2009hybrid} extend NC-HARQ by adapting the amount of FEC in real time to channel conditions. This technique increases the throughput efficiency up to 3.5 times over ARQ and 1.5 times over HARQ. Zhang et al. \cite{zhang2011dual} extend NC-HARQ by supplementing XOR retransmission coding with additional XOR operations, combining dynamically lost packets from the same receiver. Lu et al. \cite{lu2011network} study NC-HARQ in the context of wireless video broadcast. Abuzeid et al. \cite{abuzeidir} compare NC-HARQ and IR-HARQ in cooperative wireless communication systems. 

NC is further proposed in multi-hop and cooperative contexts. Fan et al. \cite{fan2009reliable}, Sun et al. \cite{sun2009cooperative} and Vien et al. \cite{vien2010network} consider a scenario where two nodes communicate with the BS with the assistance of a relay. Fan et al. \cite{fan2009reliable} introduce a NC-based cooperative multicast scheme while Sun et al. \cite{sun2009cooperative} discuss cooperative HARQ based on NC (C-HARQ-NC). Vien et al. \cite{vien2010network} investigate ARQ based on NC for two-way wireless relay networks. Recently, Vien et al. also discuss NC-based block ARQ (BACK) for wireless relay networks \cite{vien2011network}. Hong et al. \cite{hong2010network} propose NC-HARQ for mobile relay systems.

In this paper, we are interested in the use of NC in a WiMAX setting. Past work in this area includes contributions by Jin et al. \cite{jin2008random,jin2008adaptive} and Yazdi et al. \cite{yazdi2009optimum}, where NC is considered in conjunction with ARQ and/or HARQ in WiMAX. Jin et al. \cite{jin2008random} introduce MRNC and report a 10\% gain in throughput over HARQ in single-hop transmissions. The adaptive extension of MRNC \cite{jin2008adaptive} outperforms regular MRNC by 28.4\% and HARQ by 57.7\% in terms of throughput. Adaptive MRNC uses the channel state information feedback to adjust dynamically packet size according to current channel conditions. Yazdi et al. \cite{yazdi2009optimum} extend MRNC to restrict the number of retransmissions to an upper bound which is important for delay sensitive applications.

\begin{figure}[!tbp]
\centering
\includegraphics[width=\columnwidth]{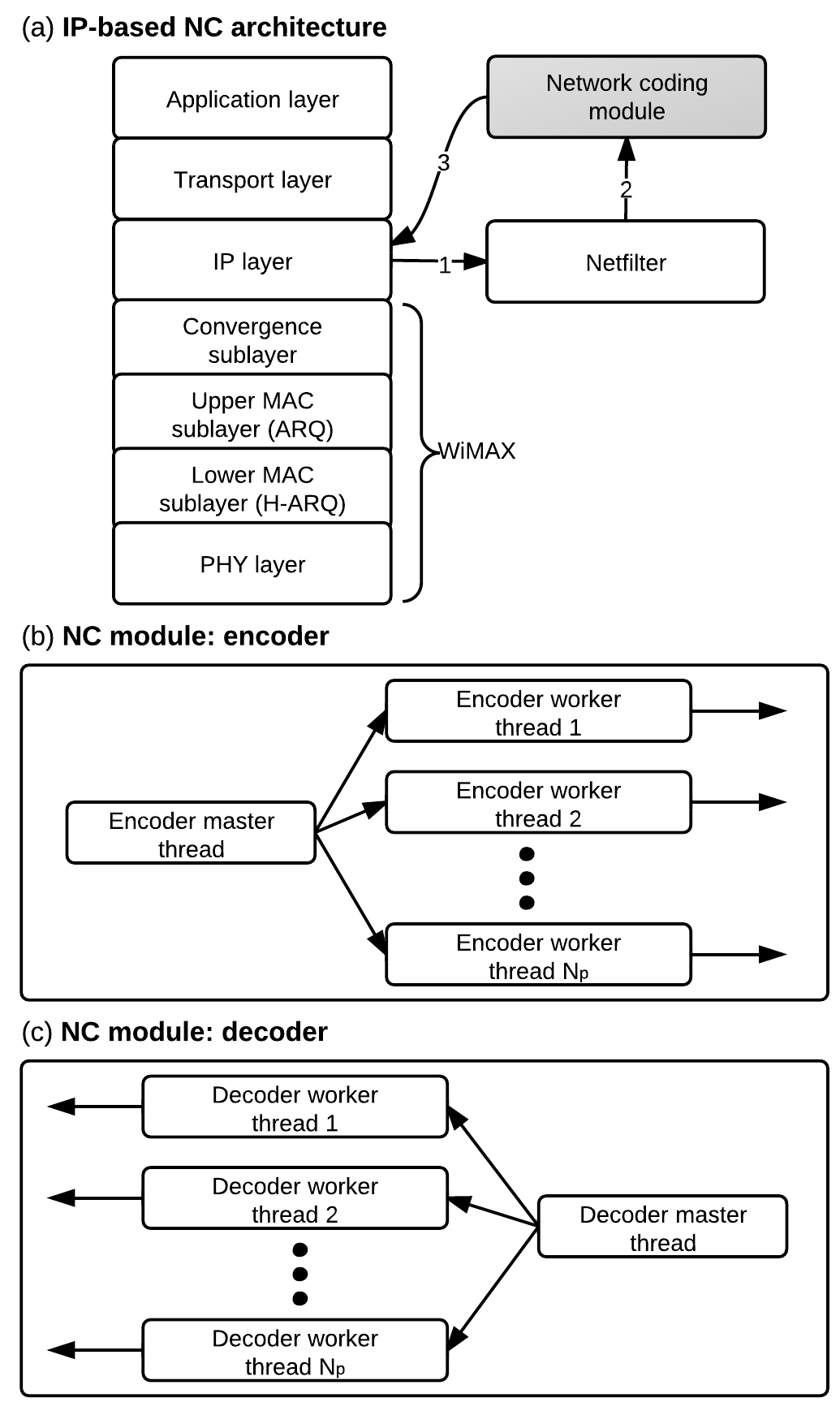}
\caption{(a) \textbf{IP-based NC architecture},  at the BS for encoding and at the SS for decoding: 1) IP packets are intercepted. 2) Netfilter copies and forwards IP packets to the NC module. 3) Processed packets are injected back.  (b) \textbf{NC module: encoder}, an encoder master thread load-balances $N_p$ encoder worker threads in a round-robin fashion. (c) \textbf{NC module: decoder}, a decoder master thread and $N_p$  decoder worker threads.}
\label{fig:overview}
\end{figure}

Despite the large number of studies covering NC as an alternative or enhancement to HARQ and ARQ mechanisms, they are limited to analysis and simulation, and none are supported by experimentation. To our knowledge, our work provides a first experimental implementation of NC as a throughput efficiency and reliability mechanism in a single-hop wireless link. After describing our NC-based design and implementation in the next sections, we compare its performance to that of ARQ and CC-HARQ in WiMAX.

\section{NC-Enhanced Architecture \label{sec:architecture}} 

\begin{table}[!b]
\centering
\caption{Design parameters for the NC Module.}
\begin{tabular}{|c|l|}
\hline
\textbf{Parameter} & \textbf{Description} \\
\hline
$N_p$ & number of concurrent encoder-decoder thread pairs\\
$L_t$ & processing length threshold of the buffer list\\
$T_i$ & processing time interval of the buffer list\\
$L_m$ & maximum length of segments\\
$N_r$ & preferred number of segments\\
$N_k$ & number of rounds of redundancy transmission\\
$N_m$ & number of redundancy packets per round\\
$T_r$ & time interval between each round\\
\hline
\end{tabular}
\label{table:parameters}
\end{table}

Our proposed NC-enabled reliability architecture is implemented in the form of an NC module at the IP layer of the network protocol stack, as shown in Fig.~\ref{fig:overview}. A Linux packet filtering framework (\textit{netfilter}) \cite{netfilter} intercepts, copies and forwards IP packets to the NC module, which in term injects processed packets back into the IP layer. 

The NC module is implemented in user-space. It acts as an encoder at the base station (BS), and as a decoder at the subscriber station (SS). At the BS, the source application, located in user-space, sends outgoing IP packets to the Operating System (OS) where the transport and IP layers are run. Netfilter intercepts those packets and sends them to the encoder NC module in user-space. The encoder returns coded IP packets to the OS. Coded IP packets then traverse the WiMAX stack, passing through the Convergence Sublayer (CS), the upper and lower MAC sublayers and the PHY layer. At the SS, netfilter intercepts the incoming coded IP packets handed from WiMAX to the OS and delivers them to the decoder NC module in user-space. The decoder sends decoded packets back to the OS, where they are forwarded to the destination application. In the NC-enhanced architecture, ARQ and HARQ, run from the upper and lower MAC sublayer respectively, are switched off.

Table~\ref{table:parameters} lists the key design parameters for the proposed NC module, while Table~\ref{table:variables} lists variables derived from these module parameters. Exact definitions of these parameters and variables will be provided in subsequent subsections, where their importance will also be discussed.
\begin{table}[!t]
\centering
\caption{Derived variables for the NC Module.}
\begin{tabular}{|c|l|}
\hline
\textbf{Variables} & \textbf{Description} \\
\hline
$L_s$ & calculated segment length\\
$N_s$ & calculated number of segments\\
$L_o$ & total length of an outgoing IP packet\\
$L_b$ & coding block length\\
$L_p$ & length of the current IP packet (temporary)\\
\hline
\end{tabular}
\label{table:variables}
\end{table}
In the rest of this Section, we describe details of the NC module implementation, starting by defining the encoder and decoder processes given in Figures~\ref{fig:overview} (b) and (c).

\subsection{Thread-Based Encoder and Decoder Design}\label{sec:processes}

\begin{figure}[!hbp]
\centering
\includegraphics[width=\columnwidth]{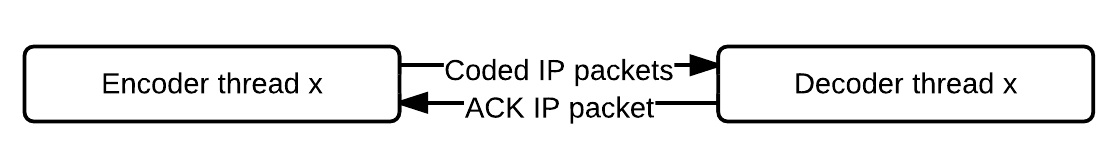}
\caption{Pair of encoder-decoder worker threads, exchanging coded IP packets and an ACK packet.}
\label{fig:pair}
\end{figure}

The NC module uses a flexible thread-based design, where parallel encoding-decoding instances are generated to process packets concurrently. Systematic intra-session RLNC is applied. The encoder and decoder processes each have a master thread and $N_p$ worker threads, as shown in Fig.~\ref{fig:overview} (b) and (c). Each encoder-decoder thread pair operates independently from other pairs and is identified by a unique Thread ID (TID). The encoder master thread load-balances encoder worker threads by distributing incoming packets in a round-robin fashion. The decoder master thread dispatches incoming coded IP packets from encoder worker threads to the corresponding decoder worker threads according to their TID. Each worker thread in the encoder process matches a unique worker thread in the decoder process, as shown in Fig.~\ref{fig:pair}. Next, we explain in more details, the encoding, decoding, and feedback mechanisms in a encoder-decoder thread pair.

\subsubsection{Encoding Mechanism}\label{subsec:encodermechanism}
 
\begin{algorithm}[!bp]
\caption{Determine when the master thread sends the buffer list to the next worker thread. $L_b$ is the length of the buffer list. $T_i$ is the time threshold to concatenate the buffer list. $L_t$ is the maximum length of the buffer list. $L_p$ is the length of the current incoming IP packet. }
\label{alg:concat}
\begin{algorithmic}[1]
   \State Initialize timer $T$
   \State Initialize length $L_b$ of buffer list 
   \While{$T < T_i$ \textbf{and} $L_b < L_t$}
   	\State Receive new packet with length $L_p$
   	\State $L_b \gets L_b + L_p$
   \EndWhile
   \State Transfer buffer list to next worker thread
\end{algorithmic}
\end{algorithm}

\begin{figure}[!tbp]
\centering
\includegraphics[width=\columnwidth]{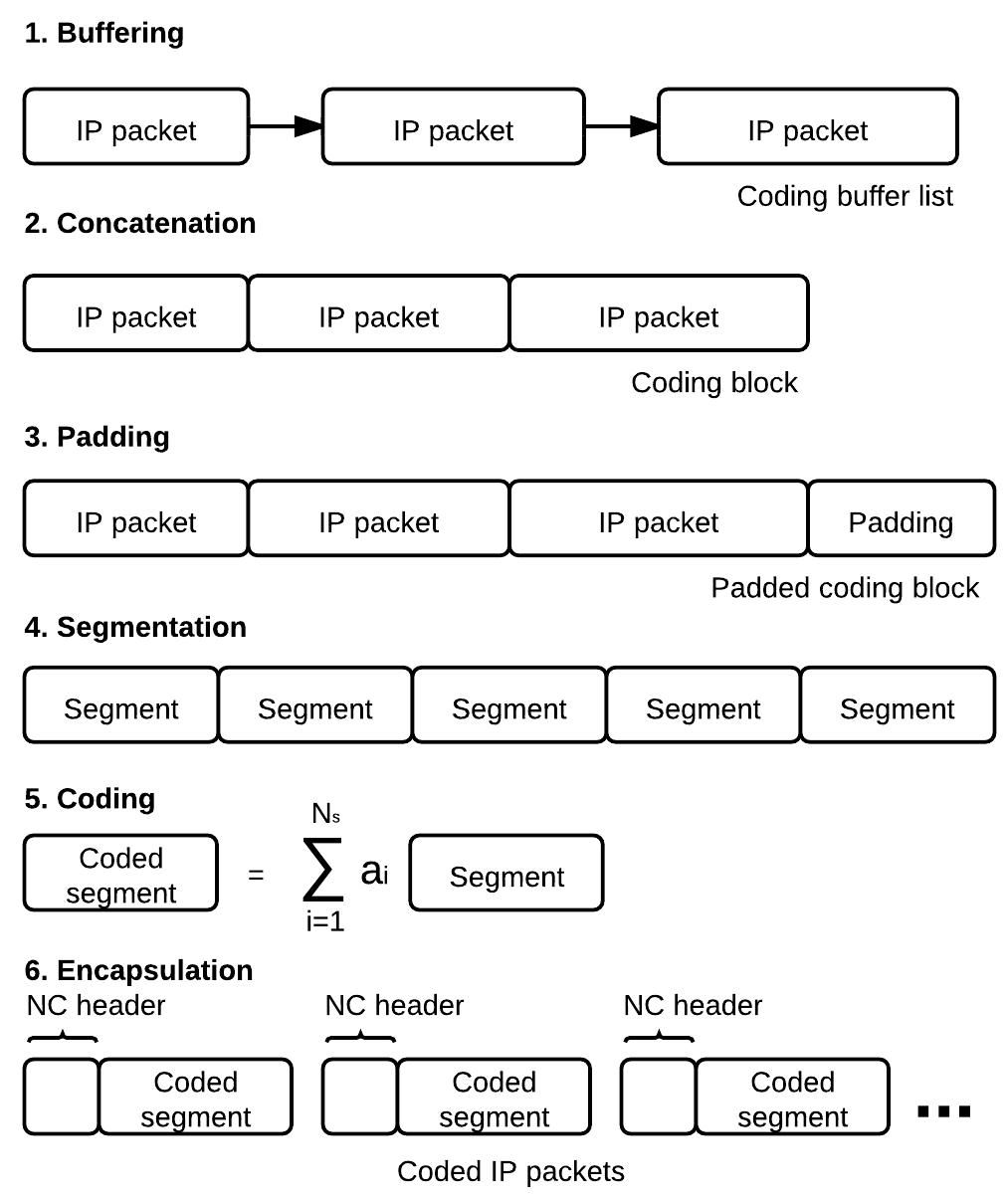}
\caption{Successive steps of the encoder: 1) Incoming IP packets are buffered at the master thread, forming a coding buffer list. Alg.~\ref{alg:concat} determines when the buffer list is handed to a worker thread. 2) At each worker thread, the list is concatenated into a coding block. 3) Alg.~\ref{alg:segment} determines the number of segments ($N_s$) and segment length ($L_s$), and byte padding is added. 4) The block is segmented. 5) Algorithm~\ref{alg:code} encodes the segments. 6) Coded segments are encapsulated into coded IP packets.}
\label{fig:encode}
\end{figure}
Fig.~\ref{fig:encode} illustrates the encoding mechanism. Incoming IP packets are first buffered at the master thread, and stored successively as a \textit{buffer list}. The master thread uses Alg.~\ref{alg:concat} to determine when the buffer list is handed to the next worker thread. At a worker thread, the buffer list is concatenated into a coding block. Next, the block is to be divided into \textit{segments}, the basic unit of operation for the NC module. The number of segments ($N_s$) and segment length ($L_s$) are calculated according to Alg.~\ref{alg:segment}. Byte padding is applied so that the padded block is a multiple of $N_s$. The block is then segmented and the resulting segments are coded according to Alg.~\ref{alg:code}. Finally, the encapsulation step adds a coding header, thus producing coded IP packets from the generated segments. Therefore, in our architecture, although segment size is constant for each block, it varies among blocks, depending on the traffic intensity.

\begin{algorithm}[!bp]
\caption{Determine the segment length $L_s$ and the number $N_s$ of segments, given the length $L_b$ of the coding block, the maximum length $L_m$ of segments, and the preferred number $N_r$ of segments.}
\label{alg:segment}
\begin{algorithmic}[1]
   \State $L_b \gets L_b + 1$ \Comment{1 byte for the padding boundary.}
   \State $L_s \gets \frac{L_b}{N_r}$
   \State $N_s \gets N_r$
   \While{$L_s > L_m$}
   	\State $N_s \gets N_s + 1$
	\State $L_s \gets \lceil\frac{L_b}{N_s}\rceil$
   \EndWhile
\end{algorithmic}
\end{algorithm}
\begin{algorithm}[!tbp]
\caption{Encode. $N_s$, $N_k$, $N_m$ and $T_r$ are defined in Tables~\ref{table:parameters} and \ref{table:variables}. Terminate immediately if an ACK for the same coding block is received.}
\label{alg:code}
\begin{algorithmic}[1]
   \For{$x = 1 \to N_s$} \Comment{generate systematic code first.}
   	\State generate an uncoded segment.
   \EndFor
   \While{ACK has not yet been received.}
   \For{$y = 1 \to N_k$}
   	\For{$z = 1 \to N_m$}
   		\State generate a coded segment.
	\EndFor
	\State wait for duration $T_r$ \\ \Comment{terminate if an ACK is received.}
   \EndFor
   \EndWhile
\end{algorithmic}
\end{algorithm}

More specifically, at the master thread, Alg.~\ref{alg:concat} determines when the buffer list is handed to the next worker thread by combining a timeout mechanism with a maximum size trigger for buffer list concatenation. Concatenation occurs before time interval $T_i$ or buffer length threshold $L_t$ are reached.

At each work thread, after concatenating a new buffer list, Alg.~\ref{alg:segment} determines $N_s$ and $L_s$ for the new coding block. Byte padding is added so that the padded block is a multiple of $N_s$. We use the well-known ANSI X.923 byte padding algorithm \cite{padding}. In ANSI X.923, bytes filled with zeros are appended to the data and the last byte stores the number of padded bytes. Alg.~\ref{alg:segment} first adds 1 byte to the coding block length $L_b$ for the padding boundary; it then initializes the segment length $L_s$ to $\frac{L_b}{N_r}$, and the number of segments $N_s$ to the preferred number of segments $N_r$. Next, $L_s$ and $N_s$ are adjusted to make $L_s$ less than or equal to the maximum length of segments $L_m$.

After padding, the coding block is segmented and coded according to Alg~\ref{alg:code}. We use \textit{systematic} RLNC, which has been shown to offer advantages in term of decoding delay \cite{lucani2010systematic} when compared to non-systematic network codes. With systematic coding, $N_s$ uncoded segments are first generated and sent, followed by coded segments generated with random coefficients. We refer to the uncoded segments as `systematic,' and the coded segments as `nonsystematic.' 
Up to $N_k$ rounds of $N_m$ nonsystematic segments are transmitted. An inter-round pause of duration $T_r$ is implemented to allow other threads to process their blocks. 
Although a received ACK terminates the encoding process, it is important to note that the encoder does not require ACK packets to operate. Encoding may be terminated when the maximum number $N_k$ of retransmission rounds is reached, thus protecting against inefficiencies due to ACK errors or losses. We implement RLNC in a Galois Field of size $2^8$, which is sufficient for practical applications \cite{ho:rlnc, chou2003practical}. Each coefficient is hence expressed in a single byte. 

\begin{figure}[!tbp]
\centering
\includegraphics[width=3.5in]{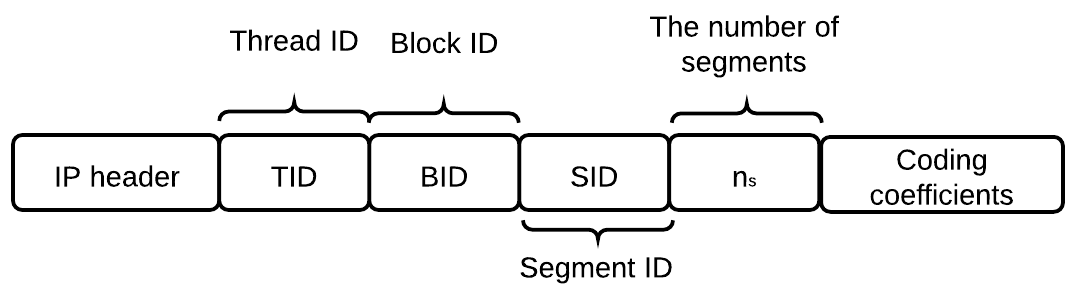}
\caption{Structure of the NC header. Possible modifications to this header are replacing the coefficients with a random seed and accounting for IP packet fragmentation  (Section~\ref{sec:ncheader}).}
\label{fig:ncheader}
\end{figure}

Coded segments are encapsulated into coded IP packets. The structure of the NC header used during encapsulation is shown in Fig.~\ref{fig:ncheader}. The NC header contains the IP header, Thread ID (TID), Block ID (BID), Segment ID (SID), number $N_s$ of segments, and coding coefficients. Segment length $L_s$ is not included because it can be derived from the packet length field in the IP header. TID identifies which thread the packet belongs to; BID identifies which block the packet belongs to within a given thread. For each thread, BID is incremented for every new coded block. SID keeps track of the number of segments generated in a particular block; it is incremented for generated packet. $N_s$ and coding coefficients are necessary for decoding, which we describe in the next section.
\subsubsection{Decoder Mechanism}\label{subsec:decodermechanism} 
At a decoder worker thread, operations described in Fig.~\ref{fig:encode} are reversed in order to recover the original IP packets. First, decapsulation strips off the NC header. For each reassembled coded block, received coded IP packets are decoded progressively using Gauss-Jordan elimination \cite{gaussjordan} based Alg.~\ref{alg:decode}. Once the block is decoded, it is unpadded and the original uncoded IP packets are separated. Note that if a packet with a different BID from the current block arrives at a decoding worker thread before the current block is decoded, the decoder will drop the current block and start decoding the new block. 
\begin{algorithm}[!bp]
\caption{Block decoding algorithm. $\mathbf{M}$ is the current coefficient matrix of incoming coded packets. $\mathbf{M}[r+1] $ refers to row $r+1$ of $\mathbf{M}$. $rank(\mathbf{M})$ is the rank of $\mathbf{M}$.}
\label{alg:decode} 
\begin{algorithmic}[1]
	\State $r \gets 0$
	\State $\mathbf{M}_{N_s\times(N_s+L_s)} \gets \mathbf{0}$
	\For{each incoming coded IP packet $N_p$}
		\State $\mathbf{M}[r+1] \gets$ coefficients and segment of $N_p$
		\State Gauss-Jordan elimination on $(r+1)\times (N_s+L_s)$ of $\mathbf{M}$
		\If{$rank(\mathbf{M}) = r+1$}
			\State $r \gets r + 1$
			\If{$r = N_s$}
				\State done decoding
			\EndIf
		\EndIf
	\EndFor
\end{algorithmic}
\end{algorithm}

\subsubsection{Feedback Mechanism}\label{subsec:feedbackmechanism} 
\begin{figure}[!tbp]
\centering
\includegraphics[width=1.8in]{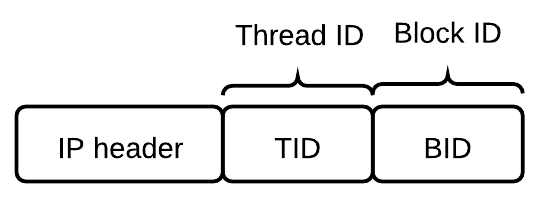}
\caption{Structure of an ACK packet}
\label{fig:ack}
\end{figure}

Once a block is decoded, the decoder worker thread sends an ACK packet to the corresponding encoder worker identified by its TID. Fig.~\ref{fig:ack} shows the structure of an ACK packet. If the encoder worker thread is still running Alg.~\ref{alg:code} on the block with the same BID as that in the ACK packet, the worker will terminate the algorithm.

\subsection{Other Design Considerations}\label{sec:others}
\subsubsection{Code Rate}\label{sec:coderate} The Code Rate (CR) of the presented design is defined as the ratio the number $N_s$ of segments, to the sum of $N_s$ and the redundancy segments:
\begin{equation}
CR \equiv \frac{N_s}{N_s+N_k\times N_m},
\end{equation} 
\noindent where $N_k$ is the number of redundancy rounds, and $N_m$ is the number of redundancy segments transmitted per round. Note that this is an upper bound on the effective code rate, as an ACK may interrupt before the transmission of $N_k$ rounds of $N_m$ redundancy segments is completed.

\subsubsection{IP Packet Fragmentation}\label{sec:implementation}
Since systematic RLNC is used, blocks that cannot be decoded can still contain useful information, as some uncoded packets may be extracted. To determine where an IP packet starts in a segment, we implement an additional two-byte field in the NC header called $start$. The $start$ field allows IP packet defragmentation at the decoder in the event of unsuccessful block decoding. 
\subsubsection{Overhead Reduction through Random Seeds}\label{sec:randomseeds}
Assuming one byte per coefficient, the total NC header length is $L_h+N_s$, where $L_h$ is the length of the NC header without coding coefficients. The NC header overhead ratio is therefore $\frac{L_h+N_s}{L_s}$, where $L_s$ is the segment length. If $N_s$ is 120, $L_h$ is 24, and $L_s$ is 1400, the overhead is 10.29\%. This overhead can be reduced in three ways: 1) by increasing $L_m$, the maximum length of segments, thus increasing $L_s$, 2) by reducing $N_s$, and 3) by sending a seed of a pseudo-random number generator instead of a coefficient vector \cite{jin2008random,jin2008adaptive,sundararajan2011network}. In this work, we implement the third option. Using random seeds, the overhead becomes $\frac{L_h+q}{L_s}$, where $q$ is the size of the seed value, typically 4 bytes. Using the previously assumed values of $L_h$ and $L_s$, the overhead is reduced to the lower value of 2\%. 

In order to support random seeds, we add new fields to the NC header: $type$ and either segment number ($segn$) or $seed$. $type$ is used to distinguish whether a packet is coded or uncoded; The parameter $segn$ is used in a systematic packet to specify the segment number; $seed$ is used in a coded packet as a random seed. We use a simple pseudo-random number generator described in Alg.~\ref{alg:pseudo}.
\begin{algorithm}[!tbp]
\caption{Gerhard's Generator: $a$ is initialized to $1$. Given the seed $a$, the function generates a pseudo-random number from $1$ to $lim$.}
\label{alg:pseudo}
\begin{algorithmic}[1]
\State $a \gets 1$
\Function{rand}{$lim$}
   \State $a \gets (a \times 32719 + 3) \mod{32749}$
   \State \Return $(a \mod{lim}) + 1$
\EndFunction 
\end{algorithmic}
\end{algorithm}

\subsubsection{Implemented NC Header}\label{sec:ncheader}
In our implementation, the 20-byte IPv4 header is augmented as follows. For all generated packets, one-byte fields representing TID, BID, SID, $N_s$, $start$ and $type$ are added. For systematic and coded packets, a one-byte $segn$ segment number or a two-byte $seed$ field is added, depending on whether the packet is systematic or coded, respectively. The IP checksum is recalculated for each generated packet according to RFC 1071 \cite{rfc1071}.
\section{Experimental Setup and Performance Metrics \label{sec:setup_and_metrics}}

The proposed architecture is implemented over a WiMAX IEEE-802.16 \cite{ieee80216ieee} downlink available through the Global Environment for Network Innovations (GENI) collaborative research framework \cite{GENI}. Four fixed downlink modulation and coding schemes (MCSs) and transmission power levels are available at the BS. For each of those PHY layer settings, 11 reliability configurations are run, including raw unreliable transmission (termed \textit{raw}) and various ARQ, HARQ and NC arrangements. For each of the reliability configurations, two transmission trials are conducted through Iperf and UFTP, respectively. HARQ and ARQ configurations default to the GENI WiMAX stations are used, partly because such default settings establish a clear baseline for comparison,  partly because we had very limited ability to adjust the PHY and MAC layer parameters. Below we provide implementation details on the PHY, MAC, and reliability configurations. 
\subsection{PHY- and MAC-Layer Settings}\label{sec:PHY_MAC_Settings}

\begin{table}[!b]
\centering
\caption{PHY- and MAC-Layer BS Configuration.}
\begin{tabular}{|l|l|}
\hline
\textbf{Parameters} & \textbf{Value} \\
\hline
PHY & OFDMA\\
Frequency & 2.59 Mhz\\
Bandwidth & 10 Mhz\\
Duplexing mode & TDD\\
Frames per second & 200 (5 ms per frame)\\
\hline
Power level & Fixed (per configuration)\\
Downlink Modulation Coding Scheme (MCS) & Fixed (per configuration)\\
Uplink Modulation Coding Scheme (MCS) & Fixed at QPSK CTC 1/2\\
\hline
HARQ\_TYPE & CC\\
HARQ\_MAX\_UL\_BURST & 1\\
HARQ\_MAX\_DL\_BURST & 1\\
HARQ\_UL\_ACK\_DELAY & 3 frames\\
HARQ\_DL\_ACK\_DELAY& 1 frame\\
HARQ\_PDU\_SN & ON\\
HARQ\_MAX\_RETRANSMISSION & 4\\
\hline
ARQ\_RETRY\_TIMEOUT & 100 ms\\
ARQ\_BLOCK\_SIZE & 256 Bytes\\
ARQ\_WINDOW\_SIZE & 1024\\
ARQ\_TX\_ACK\_DELAY & 0 ms\\
ARQ\_ACK\_PROC\_TIME & 0 ms\\
ARQ\_BLOCK\_LIFETIME & 500 ms\\
ARQ\_DLV\_ORDER & ON\\
ARQ\_RX\_PURGE\_TIMEOUT & 500 ms\\
ARQ\_SYNC\_LOSS\_TIMEOUT & 1000 ms\\
\hline
\end{tabular}
\label{table:bsparams}
\end{table}
 
At the physical layer, four fixed downlink MCSs and BS transmission power levels are available, with increasing PHY code rates and power levels. These are listed in Table \ref{table:PHY_Settings}, along with Carrier to Interference plus Noise Ratio (CINR), Received Signal Strength Indication (RSSI) and Average Tx Power, measured at the SS. 

\begin{table}[!t]
\centering
\caption{Available PHY transmission settings with resulting Carrier to Interference plus Noise Ratio (CINR), Received Signal Strength Indication (RSSI) and Average Tx Power, measured at the SS.} \label{table:PHY_Settings}
\begin{tabular}{|c|c|c|c|c|}
\hline
\multicolumn{2}{|c|}{\textbf{BS}} & \multicolumn{3}{|c|}{\textbf{SS}}\\
\cline{1-5}
\textbf{MCS} & \textbf{Tx. Power} & \textbf{CINR} & \textbf{RSSI} & \textbf{Tx. Power}\\
\hline
64 QAM CTC 1/2 & 13 dBm & 13 dB & -76 dBm & -63 dBm\\
\hline
64 QAM CTC 2/3 & 17 dBm & 17 dB & -76 dBm & -63 dBm\\
\hline
64 QAM CTC 3/4 & 18 dBm & 18 dB & -75 dBm & -63 dBm\\
\hline
64 QAM CTC 5/6 & 20 dBm & 18 dB & -73 dBm & -63 dBm\\
\hline
\end{tabular}
\end{table}
\begin{table}[!tbp] 
\centering
\caption{Effective PHY-layer data rates.}
\begin{tabular}{|l|c|c|}
\hline
 & \textbf{MCS} & \textbf{PHY rates}\\
\hline
\textbf{Uplink} & QPSK, 1/2 & 1.344 \\
\hline
\multirow{4}{*}{\textbf{Downlink}} & 64 QAM, 1/2 & 15.120 \\
&64 QAM, 2/3 & 20.160 \\
&64 QAM, 3/4 & 22.680 \\
&64 QAM, 5/6 & 25.200 \\
\hline
\end{tabular}
\label{table:phydatarate}
\end{table}

For each PHY setting, we run a number of reliability configurations involving different NC, HARQ, and ARQ settings. The implemented BS PHY-layer parameters are shown in Table~\ref{table:bsparams}. Also shown are HARQ and ARQ parameters when both are turned on.
These represent the default equipment configuration, whereby CC-HARQ is employed. The maximum number of HARQ UpLink (UL) bursts per frame is set to 1. The maximum number of HARQ DownLink (DL) bursts per frame is set to 1. The frame offset between an UL burst with respect to its UL ACK is set to 3. The frame offset of the DL ACK is set to 1. In addition, PDU SN extended sub-header reordering is enabled and the maximum number of retransmissions is set to 4.

For ARQ, we use the following default settings. The minimum time interval a transmitter will wait before retransmission of an unacknowledged ARQ block is set to 100 ms, where the interval starts at the last block transmission. ARQ block size is set to 256 bytes. The transmission window size (i.e., number of queued ARQ ACK blocks at any given time) is set to 1024. ACK processing time is set to 0. The maximum time interval an ARQ block will be managed by the transmitter ARQ state machine, once initial transmission of the block has occurred, is set to 500 ms. If transmission (or subsequent retransmission) of the block is not acknowledged by the receiver before this time limit is reached, the block is discarded. In-order delivery is enabled. The time interval the receiver will wait after successful reception of a block that does not result in advancement of ARQ\_RX\_WINDOW\_START value is set to 500 ms. Lastly, the maximum time interval ARQ\_TX\_WINDOW\_START or ARQ\_RX\_WINDOW\_START parameters can stay at the same value before declaring a loss of synchronization between transmitter and receiver is set to 1000 ms.

It is important to note that, owing to the fixed 10~Mhz channel bandwidth, the available PHY settings of Table \ref{table:PHY_Settings} yield the PHY-layer data rates shown in Table~\ref{table:phydatarate} \cite{andrews2007fundamentals}, where the data rates relevant to our experiments are highlighted.

\subsection{Reliability Configurations \label{sec:Reliability_Configurations}}
For each PHY setting, the 11 tested reliability configurations are shown in Table~\ref{table:conf}, where $N_m$ is the number of redundancy packets per round in NC. 
\begin{table}[!t]
\centering
\caption{Reliability Configurations.}
\begin{tabular}{|l|l|l|l|}
\hline
\textbf{Configuration} & \textbf{ARQ} & \textbf{HARQ} & \textbf{NC} \\
\hline
Raw & OFF & OFF  & OFF\\
HARQ & OFF & ON  & OFF\\
HARQ-ARQ & ON & ON  & OFF\\
NC-10 & OFF & OFF  & ON, $N_m=10$\\
NC-15 & OFF & OFF  & ON, $N_m=15$\\
NC-20 & OFF & OFF  & ON, $N_m=20$\\
NC-24 & OFF & OFF  & ON, $N_m=24$\\
NC-30 & OFF & OFF  & ON, $N_m=30$\\
NC-40 & OFF & OFF  & ON, $N_m=40$\\
NC-60 & OFF & OFF  & ON, $N_m=60$\\
NC-120 & OFF & OFF  & ON, $N_m=120$\\
\hline
\end{tabular}
\label{table:conf}
\end{table}

We set NC parameters to the simple configurations summarized in Table~\ref{table:ncappparams}. A single thread ($N_p=1$) is implemented, and a single redundancy round ($N_k=1$) of $N_m$ packets is transmitted immediately ($T_r=0$) after the block. The processing length threshold ($L_t$) and processing time interval ($T_i$) of the buffer list are set to 22400~bytes and 1~s, respectively. Finally, The maximum segment length ($L_m$) and preferred number of segments (i.e., initial block size, $N_r$) are set to 1400~bytes and 120 segments, respectively. 

\begin{table}[!b]
\centering
\caption{NC parameters.}
\begin{tabular}{|c|l|}
\hline
\textbf{Parameters} & \textbf{Value} \\
\hline
$N_p$ & 1\\
$N_k$ & 1\\
$T_r$ & 0 ns\\
$T_i$ & 1~s\\
$L_t$ & 22400 bytes\\
$L_m$ & 1400 bytes\\
$N_r$ & 120\\
$N_m$ & Follows NC configuration index\\

\hline
\end{tabular}
\label{table:ncappparams}
\end{table}

For each NC configuration, the approximate NC Code Rate (CR), defined in Section \ref{sec:coderate}, is calculated and shown in Table~\ref{table:coderate}. For instance, in NC-10, the block size $N_r$ is 120 and $N_m$ is 10. Therefore, 130 packets are sent per block, achieving a CR of 12/13. In theory, for best performance, the CR will match the raw throughput percentage.

\begin{table}[!t]
\centering
\caption{Code Rate (CR) for NC Configurations.}
\begin{tabular}{|c|c|}
\hline
\textbf{NC Configuration} & \textbf{Code Rate (CR)}\\
\hline
NC-10 & $12/13 = 0.92$\\
NC-15 & $8/9 = 0.89$\\
NC-20 & $6/7 = 0.86$\\
NC-24 & $5/6 = 0.83$\\
NC-30 & $4/5 = 0.80$\\
NC-40 & $3/4 = 0.75$\\
NC-60 & $2/3 = 0.67$\\
NC-120 & $1/2 = 0.50$\\
\hline
\end{tabular}
\label{table:coderate}
\end{table}

\subsection{Transmission Trials} 

In our experiments, UDP traffic is used as an emulation of real-time traffic. For each PHY setting and reliability configuration, two transmission trials are conducted through Iperf and UFTP, respectively. Iperf~\cite{iperf} is an application-layer network performance tool capable of creating UDP streams for throughput measurements, UFTP~\cite{uftp} is a UDP-based FTP application. Both our measurement tools thus deploy UDP as the underlying transport protocol. In both the Iperf and UFTP trials, an application-layer load of 6~Mbps is offered at a fixed 1400-byte packet-size. Each individual Iperf trial is terminated after a fixed duration of 60~seconds, whereas the UFTP transmissions are run until a 50~MByte file is successfully transferred. Note that the offered load of 6 Mbps is well below the effective downlink PHY-layer data rates shown in Table \ref{table:phydatarate}. Note that the measured losses are observed at the application layer (through Iperf), while lower-layer statistics are not available in our experiments.

\subsection{Performance Metrics \label{sec:Performance_Metrics}}
For each reliability configuration, we report the following performance metrics.

\subsubsection{Downlink Iperf loss percentage}
the percentage of packets lost over the total number of packets sent by Iperf, at the application layer, over the duration of the experiment.  

\subsubsection{Downlink Iperf throughput, loss and redundancy bandwidth}
the throughput is the number of packets successfully received by Iperf over the duration of the experiment, at the application layer. Two related values are the bandwidth loss and the redundancy bandwidth. The loss is calculated by subtracting the throughput from the offered load. The redundancy bandwidth is the additional bandwidth used beyond the offered load for the propose of redundancy. In the raw case, the redundancy bandwidth is 0. For HARQ and HARQ-ARQ, since performance measurements are not available within the WiMAX stack, we assume a best-case scenario where redundancy bandwidth is also 0. For NC, we simply approximate the redundancy bandwidth as $$\frac{N_m}{N_r}\times o,$$ where $N_m$ is the number of redundancy packets per round, $N_r$ is the preferred number of segments and $o$ is the offered load. Note that for exact calculations, the computed number of segments $N_s$ should replace $N_r$, and the actual redundancy bandwidth should include the NC header overhead.

\subsubsection{Downlink Iperf Throughput to Loss plus Redundancy Ratio (TLR)}
TLR is calculated as $$TLR \equiv \frac{T}{L+R},$$
where $T$ is the throughput, $L$ is the lost bandwidth and $R$ is the redundancy bandwidth. An efficient scheme should give high throughput while keeping lost and redundancy bandwidth low. Thus, TLR is a measure of efficiency.

\subsubsection{Downlink UFTP file transfer delay}\label{UFTP_delay}
in UFTP, a file is divided and packetized into UDP packets of a specified length \cite{uftp}. The transmitter sends the packets; the receiver responds with NACKs for missing packets; the transmitter then resends the missing packets. File transfer is completed when the transmitter receives no NACKs from the receiver.

\section{Results and Discussion}\label{sec:results}
In this section, we first present the results of all reliability configurations for one of the PHY settings. We then summarize and discuss the results for all PHY settings.

\subsection{Case-Study: 64 QAM CTC 5/6 at 20 dBm}\label{64qam2120}
The MCS and power level considered in this section is 64 QAM CTC 5/6 at 20 dBm. As shown in Table~\ref{table:PHY_Settings}, this PHY setting yields SS-measured CINR, RSSI and Average Tx Power values of 18~dB, -73~dBm and -63~dBm, respectively. 
\begin{figure}[t!]
\centering
\includegraphics[width=\columnwidth]{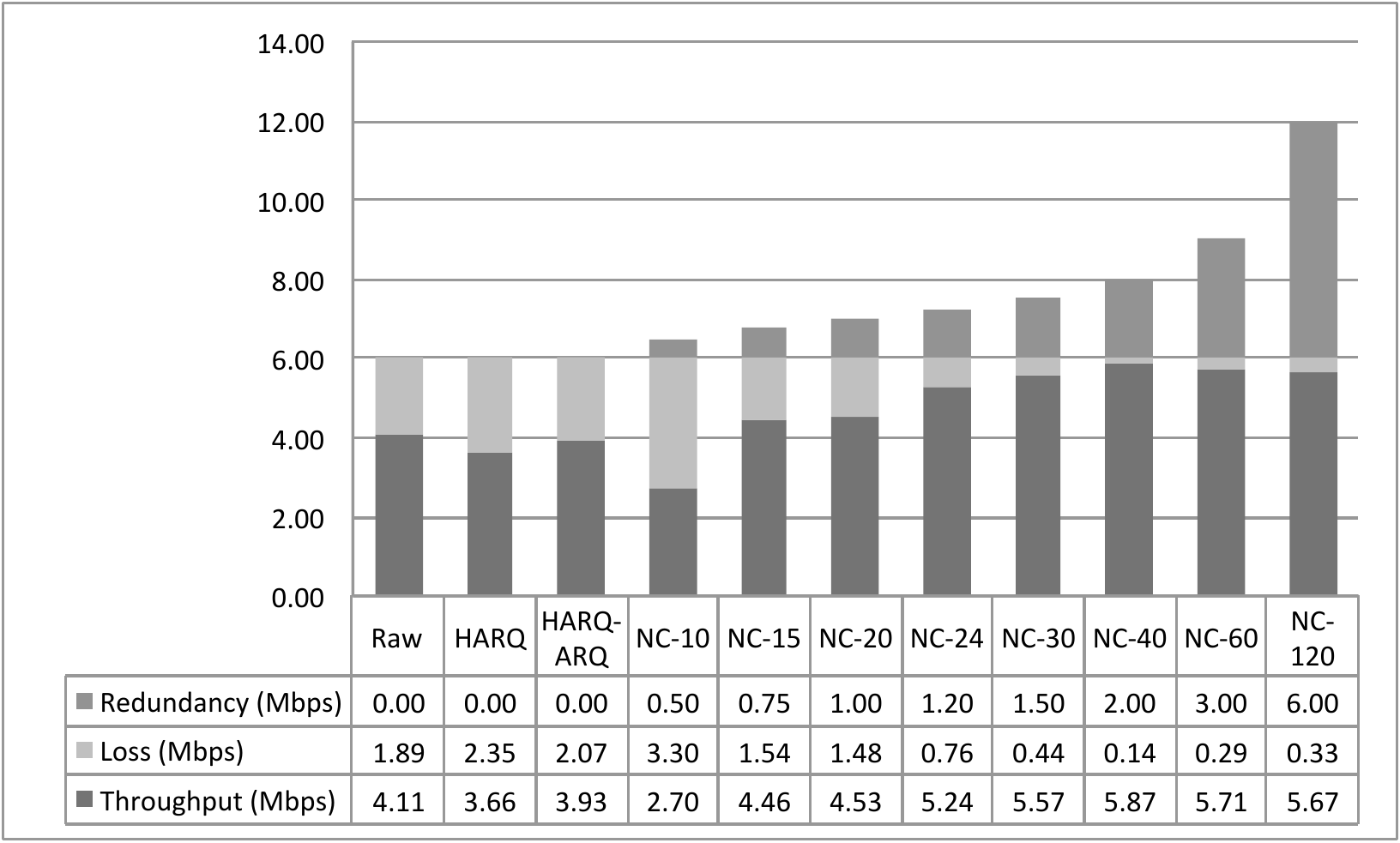}
\caption{Downlink throughput, lost bandwidth and redundancy bandwidth for 64 QAM CTC 5/6 at 20 dBm under an offered load of 6 Mbps (Iperf measurements averaged over 60~s).}
\label{fig:t2120}
\end{figure}
\begin{figure}[t!] 
\centering
\includegraphics[width=\columnwidth]{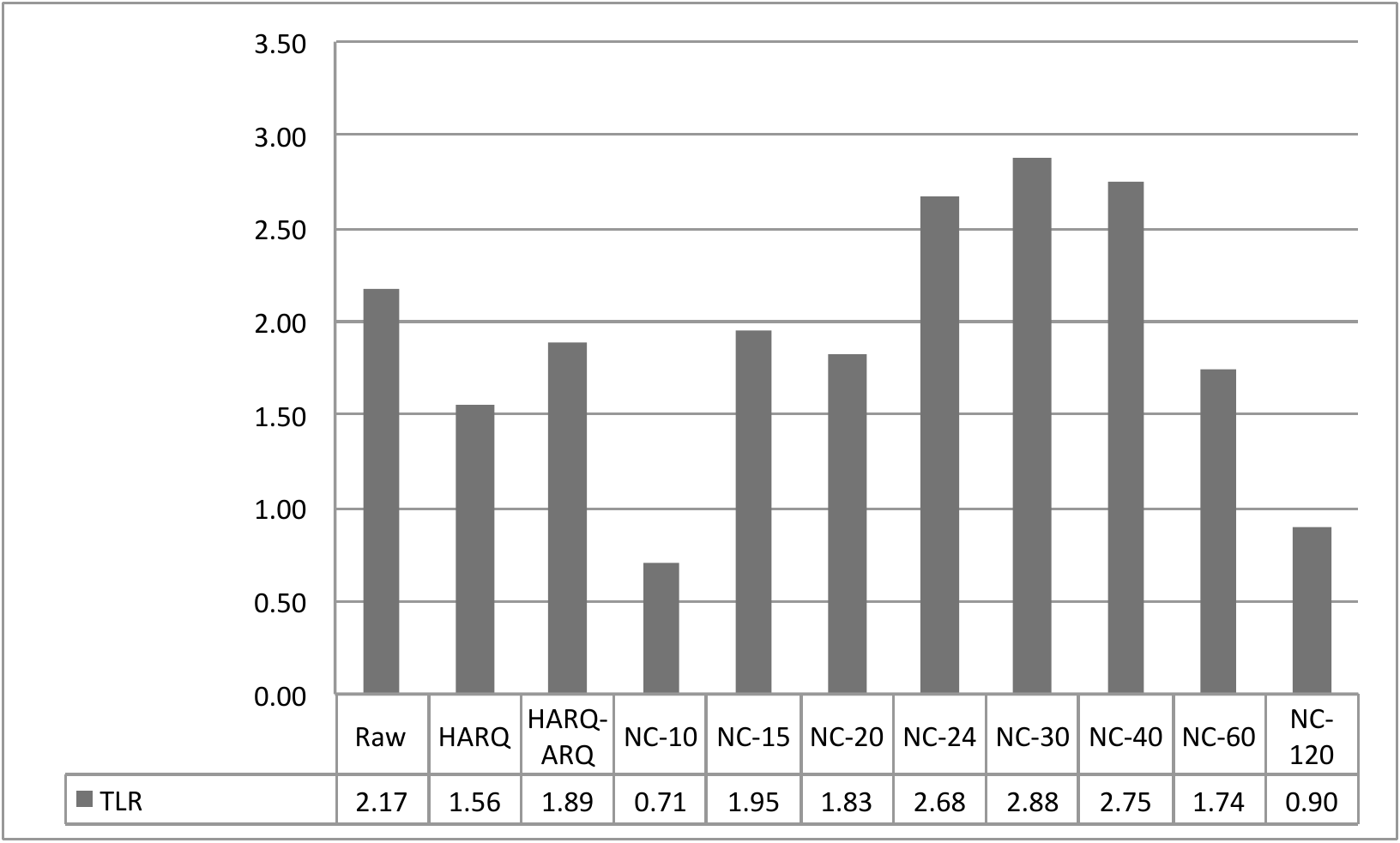}
\caption{Downlink Throughput to Loss plus Redundancy Ratio (TLR) for 64 QAM CTC 5/6 at 20 dBm under an offered load of 6 Mbps (Iperf measurements averaged over 60~s).}
\label{fig:lelr2120}
\end{figure}

The histogram of Fig.~\ref{fig:t2120} combines throughput, lost bandwidth and redundancy bandwidth results for each of the reliability configurations, listed on the horizontal scale. Throughput is represented in dark-gray at the base of the histogram columns, whereas loss is represented in light-gray above throughput. Redundancy bandwidth is shown at the top of the NC columns. No redundancy bandwidth is shown for HARQ and ARQ since relevant MAC- and PHY-layer information is unavailable. Quantities represented in Fig.~\ref{fig:t2120} are averaged over a duration of 60~s through Iperf, at the application layer.

Observe that some of the NC configurations yield significant throughput increases compared to the raw, HARQ and HARQ-ARQ configurations. The highest throughput is achieved by NC-40 and is at least 40\% above any of the non-NC configurations. Throughput of the NC configurations increases steadily with the redundancy level until it reaches 90\% of the offered load (6~Mbps). As expected, packet losses also decrease for rising values of $N_m$. Among all the configurations tested in this PHY setting, NC-10 exhibits the highest packet loss while NC-40 has the lowest. Moreover, the reduction in packet losses of HARQ-ARQ compared to HARQ (~12\%) illustrates potential benefits of ARQ under this PHY setting.

Fig.~\ref{fig:t2120} shows that all the NC configurations except NC-10 improve throughput and reduce loss compared to the raw case. In contrast, the throughput and loss performance achieved by HARQ and HARQ-ARQ are slightly below raw. We provide some possible reapons for this observation in Section~\ref{subsec:discussion}. Also keep in mind that the raw throughput is raw unreliable throughput, whereas HARQ and ARQ throughput represent reliable in-order packet flows.

Fig.~\ref{fig:lelr2120} shows TLR measured through Iperf. For the NC configurations, TLR values exhibit a maximum around $N_m=30$, with levels increasing for $N_m\le30$ and decreasing for $N_m\ge30$. Hence, an optimal redundancy value that achieves the highest throughput using minimal redundancy (redundancy bandwidth) exists at around $N_m=30$. Note that although high throughput levels are achieved for all $N_m \le 40$ (see Fig.~\ref{fig:t2120}), none are as efficient as NC-30.  

The lowest throughput and TLR occurs at $N_m=10$. At low redundancy values, $N_m$ is not sufficient to compensate for channel losses. As a consequence, blocks get discarded at the transmitter before a sufficient number of coded segments is received. At excessive redundancy values, on the other hand, the increase in overhead consumes valuable channel resources from data transmission, reducing efficiency and throughput. The described tradeoff is clear in the TLR profile of Fig.~\ref{fig:lelr2120}. Note that the configurations with the highest TLR, NC with $24\le N_m \le 40$, outperform the raw configuration, which transmits no redundancy packets.

\begin{figure}[!t]
\centering
\includegraphics[width=\columnwidth]{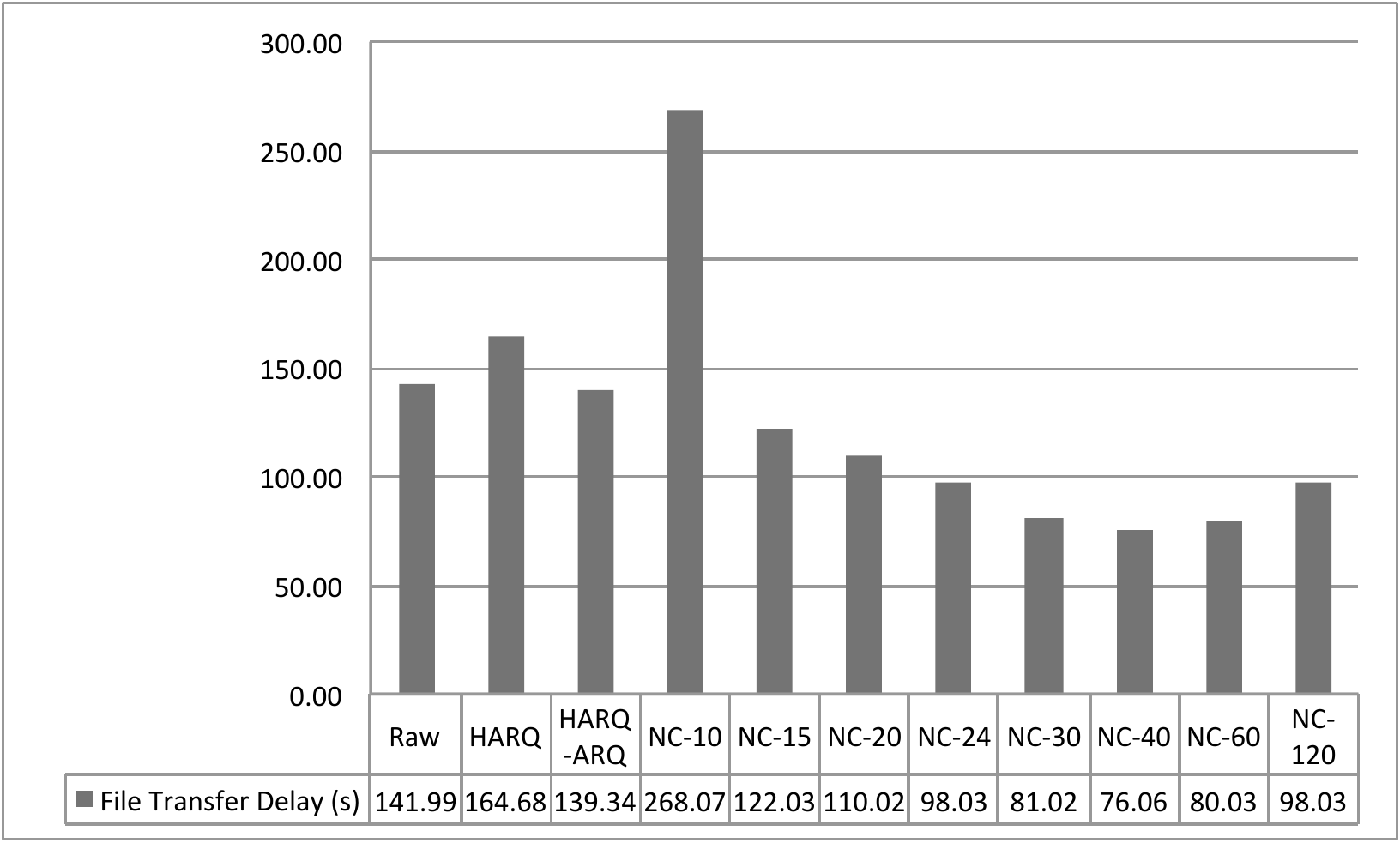}
\caption{Downlink file transfer delay for 64 QAM CTC 5/6 at 20 dBm over an offered load of 6 Mbps (UFTP measurements for 50~MB file).}
\label{fig:ete2120}
\end{figure}

Fig.~\ref{fig:ete2120} shows the file transfer delay for all tested configurations at this PHY setting. The delay profile is the inverse of the TLR profile, with a minimum around NC-40. With a 46\% reduction from that of raw, NC-40 provides the best delay performance. As mentioned in Section \ref{UFTP_delay}, UFTP is capable of completing the file transfer over an unreliable link (e.g., raw and NC-10 configurations) by resorting to its own reliability mechanism at the application layer.

\subsection{Summary of Results}\label{subsec:summary}
\begin{figure}[t!]
\centering
\includegraphics[width=\columnwidth]{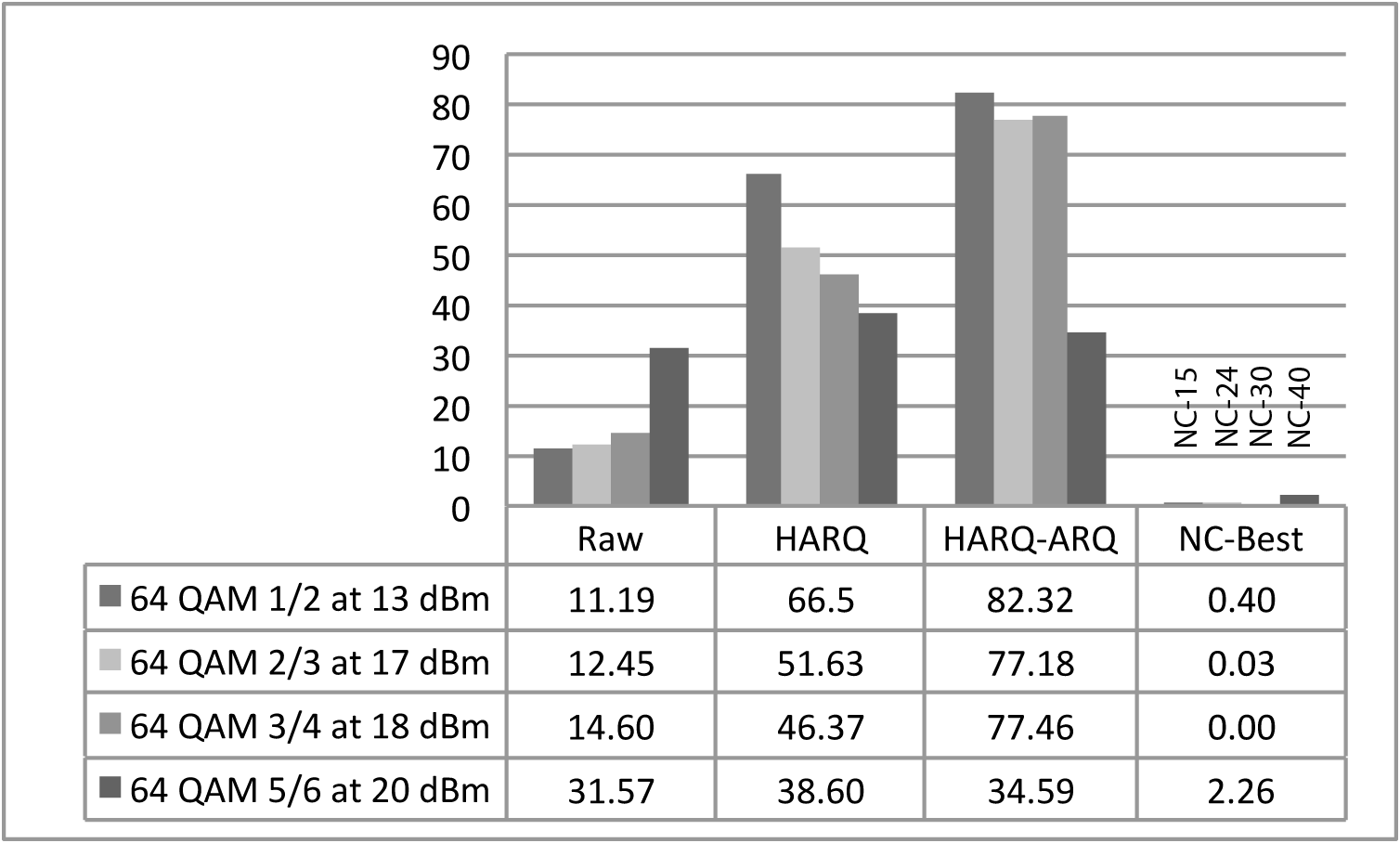}
\caption{Average Downlink Loss (\%) Comparison under an offered load of 6~Mbps (Iperf measurements over 60~s)}
\label{fig:lall}
\end{figure}

In this section, we compare the results of the raw, HARQ and HARQ-ARQ configurations with the best NC configuration at all four PHY settings. The best NC configuration, termed \textit{NC-Best} in the result figures, is the one yielding the highest performance for any given measured metric.

Fig.~\ref{fig:lall} shows the average downlink loss percentages for all four PHY settings. As expected, as code rate increases, the raw loss percentage increases. However, the loss percentage for HARQ and HARQ-ARQ are higher for lower code rates. The use of ARQ, in particular, increases losses significantly (15\%--25\%) under the three PHY settings with the lower rates. In contrast, the best NC configuration keeps the loss percentage close to 0\% for all PHY settings.

\begin{figure}[t!]
\centering
\includegraphics[width=\columnwidth]{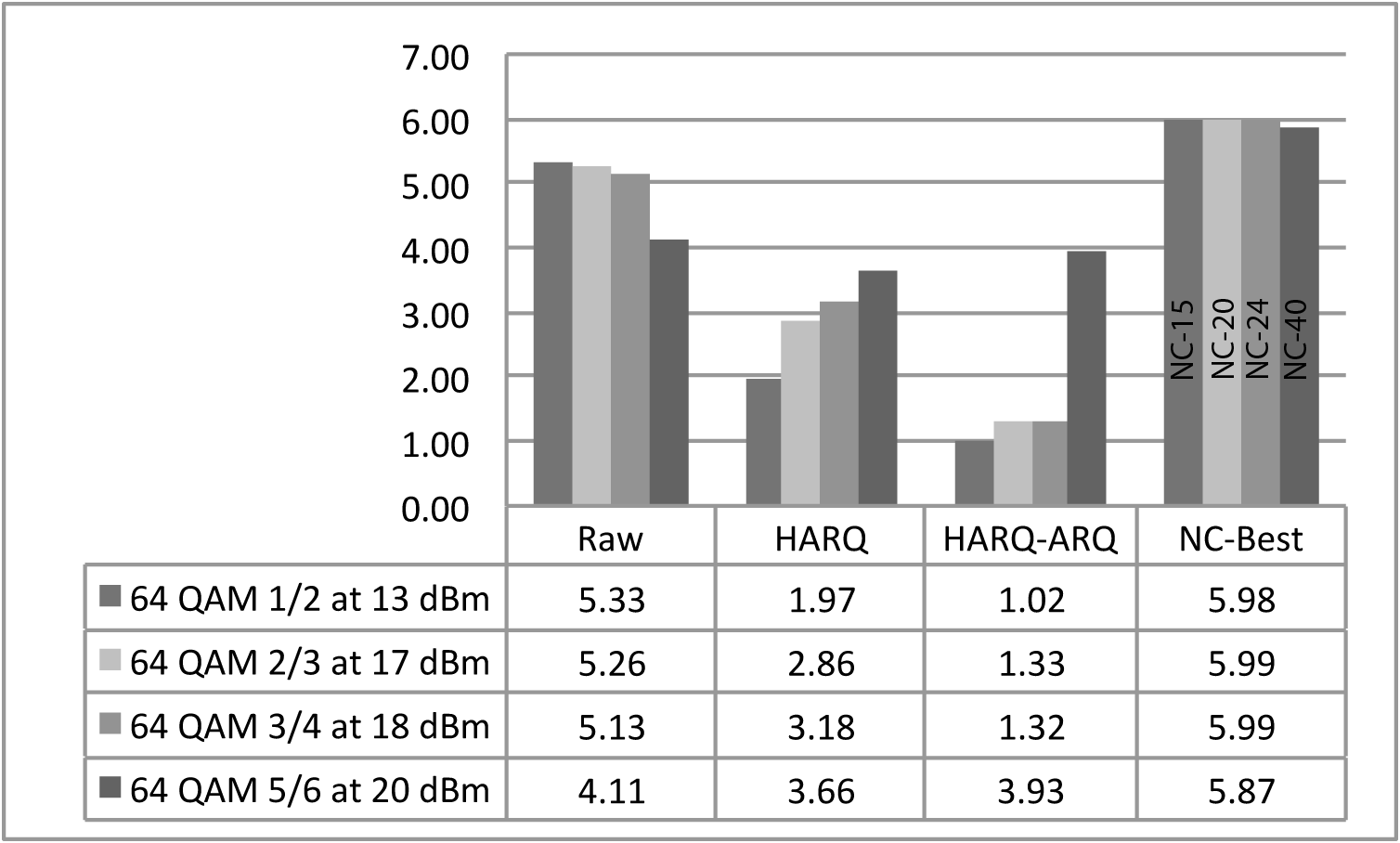}
\caption{Downlink Throughput (Mbps) Comparison under an offered load of 6~Mbps (Iperf measurements over 60~s)}
\label{fig:tall}
\end{figure}

Fig.~\ref{fig:tall} shows the average downlink throughput for all four PHY settings under the offered load of 6~Mbps. These results mirror the loss percentage results: as PHY code rate increases, raw throughput decreases, whereas HARQ and HARQ-ARQ throughputs increase. The inefficiency of HARQ/ARQ at low PHY code rates may be due to the lower data rate available for downlink retransmissions, as shown in Table~\ref {table:phydatarate}. The best NC configuration keeps throughput close to the full offered load of 6~Mbps. Therefore, in addition to introducing a high level of reliability, NC is capable of multiplying the raw throughput by up to 1.4 times. More significantly, it multiplies the throughput of HARQ and HARQ-ARQ by up to 3.0 and 5.9, respectively.

\begin{figure}[t!]
\centering
\includegraphics[width=\columnwidth]{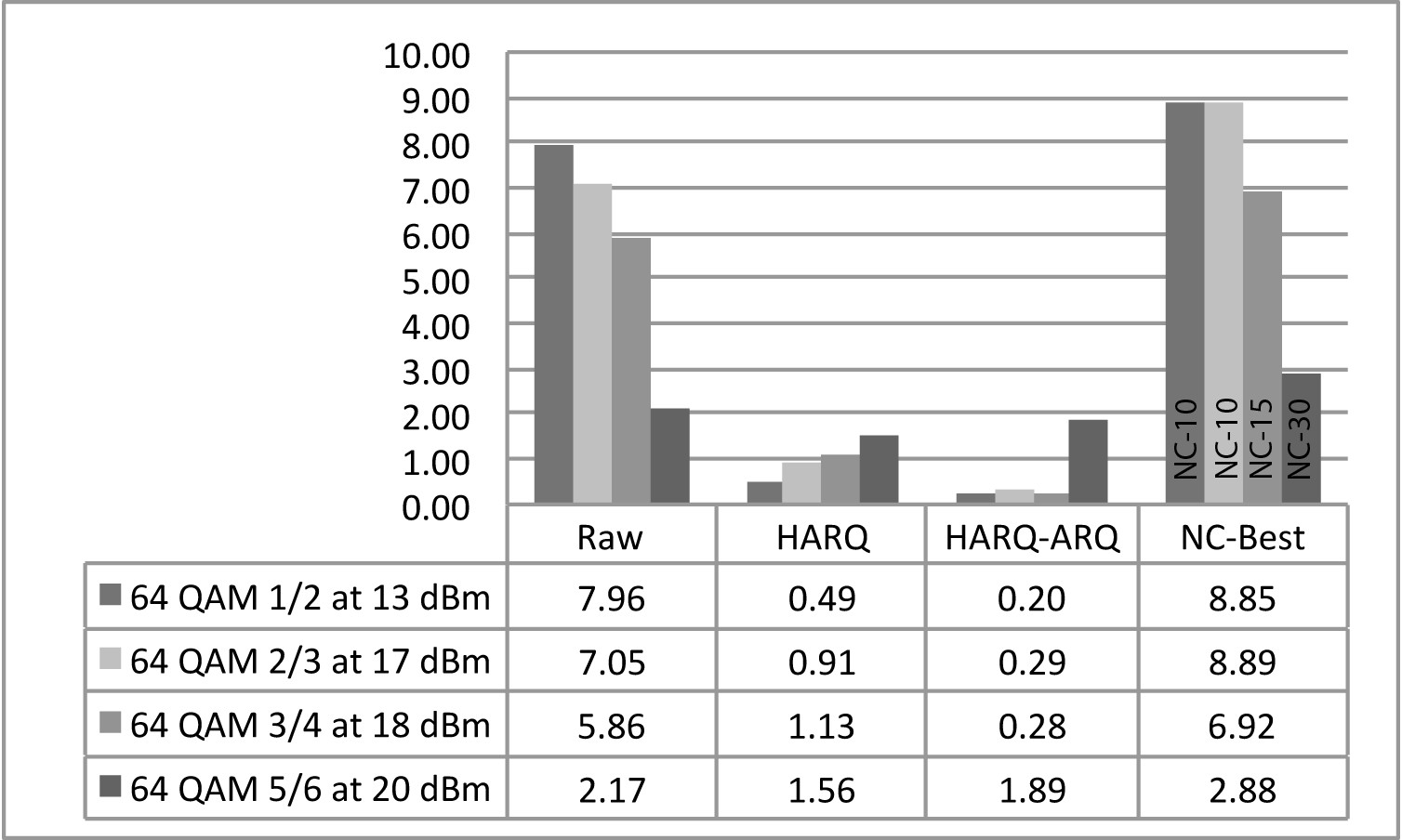}
\caption{Throughput to Loss plus Redundancy Ratio (TLR) Comparison under an offered load of 6~Mbps (Iperf -- 60~s).}
\label{fig:tlerall}
\end{figure}

Fig.~\ref{fig:tlerall} depicts the TLR for all four PHY settings under the offered load of 6~Mbps. As code rate increases, losses grow, leading to decreasing raw TLR levels. The NC configurations exhibit a similar decreasing profile. Although NC removes losses seen in the raw configuration almost entirely through redundancy, it remains more efficient than raw for all PHY settings. As in the case of loss levels, TLR levels increase with higher PHY code rates for HARQ and HARQ-ARQ. Despite ignoring any potential redundancy bandwidth in HARQ and HARQ-ARQ, those remain less efficient than NC.
\begin{figure}[t!]
\centering
\includegraphics[width=\columnwidth]{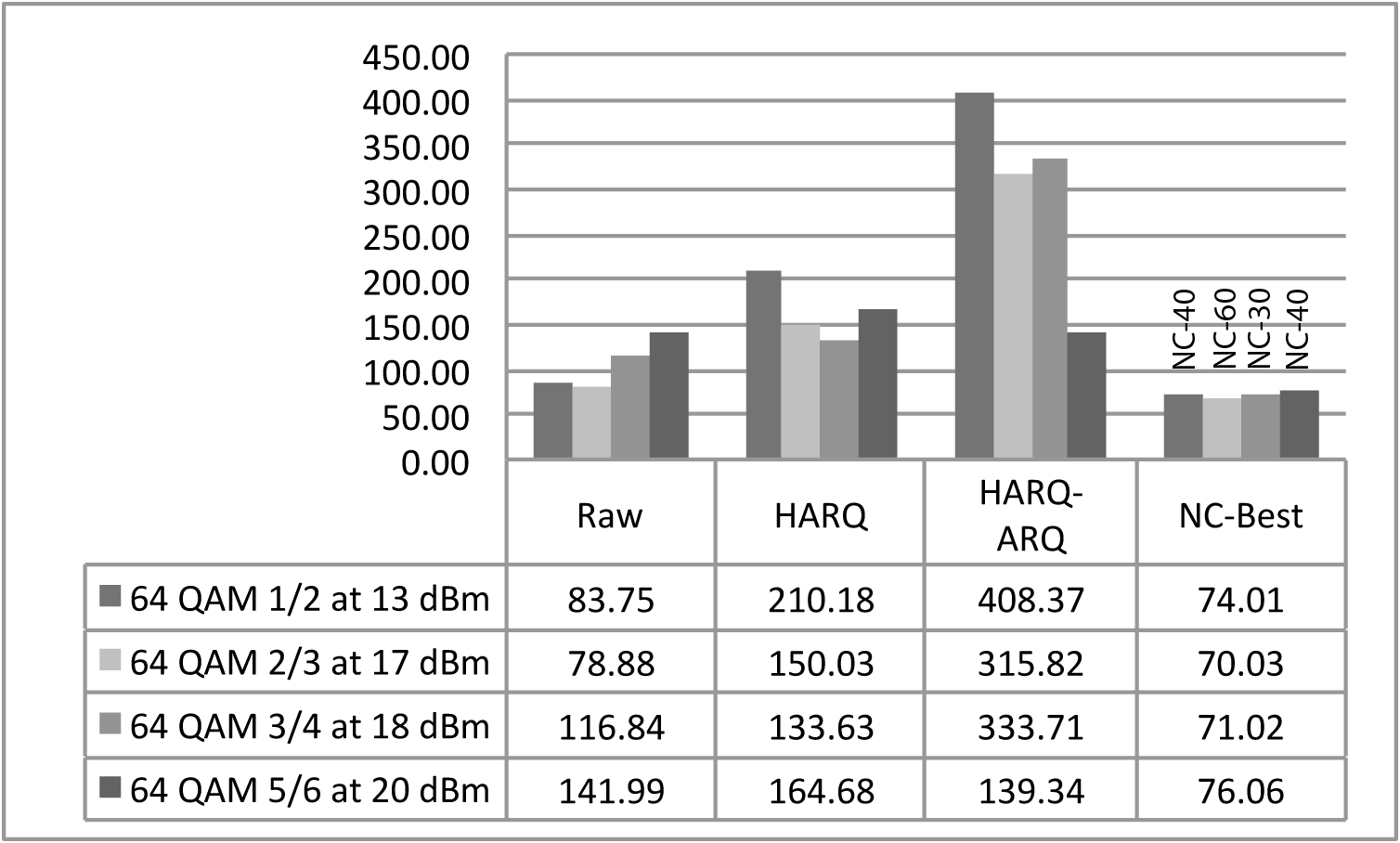}
\caption{Downlink file transfer delay (s) Comparison under an offered load of 6~Mbps (UFTP -- 50~MB file).}
\label{fig:eteall}
\end{figure}

Fig.~\ref{fig:eteall} shows the downlink file transfer delay (s) for all PHY settings. Observe that as PHY code rate increases, raw delay tends to increase. It is important to note that the delay figures demonstrated in this work apply to best effort (BE) traffic flows. In HARQ and HARQ-ARQ, the delay tends to decrease, thus confirming the higher efficiency of those reliability configurations at higher PHY code rates and CINR levels. Owing to its lower packet losses, NC maintains the lowest transfer delay of all the tested configurations, with a delay around 70~s. NC reduces the file transfer delay by 1.9 times compared to raw, 2.8 times compared to HARQ and 5.5 times compared to HARQ-ARQ.

\subsection{Discussion \label{subsec:discussion}}

The trend across different PHY settings is consistent: NC configurations use the redundancy bandwidth to increase throughput and reduce losses significantly. In contrast, HARQ and HARQ-ARQ reduce throughput and increase losses, particularly at lower PHY code rates. The loss percentage graph of Fig.~\ref{fig:lall} shows that NC works well as a packet erasure code. 

\begin{figure}[t!]
\centering
\includegraphics[width=\columnwidth]{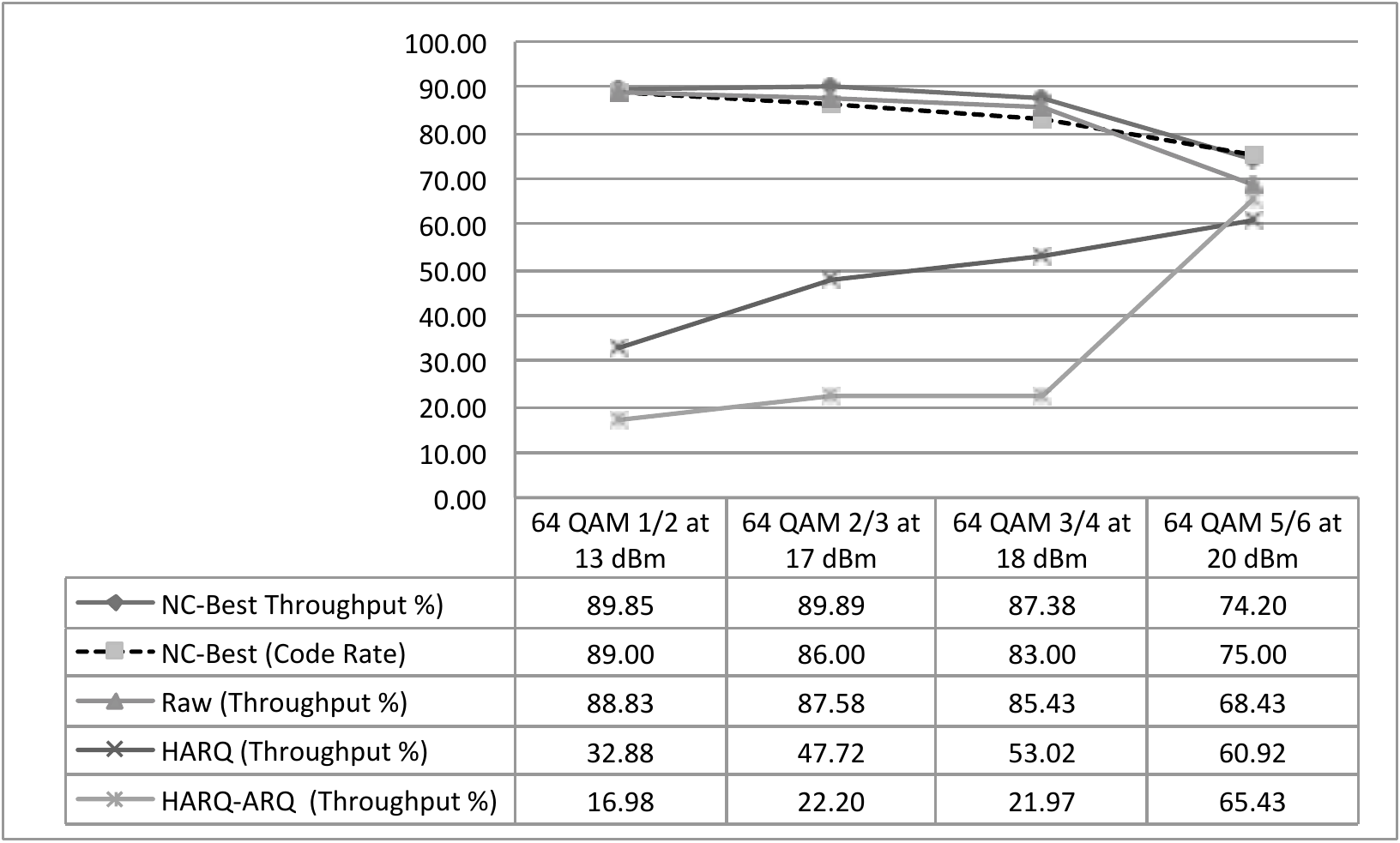}
\caption{The throughput percentage of Raw, HARQ and HARQ-ARQ, NC-Best compared to the CR of the NC-Best (the NC configuration with the highest throughput).}
\label{fig:comparecoderate}
\end{figure}

The amount of loss reduction and throughput gains of NC configurations depend on the number $N_m$ of redundancy packets per round. First, from the results shown in Section \ref{64qam2120}, we can see that a large $N_m$ may not be necessary, while a small $N_m$ may not be sufficient. When $N_m$ is too small, most coded blocks cannot be recovered, incurring additional loss, reducing throughput and increasing file transfer delay. When $N_m$ is too large, redundant packets become overheads, leading to possible buffer overflows. 
Intuitively, the optimal $N_m$ should be at a level that makes the resulting NC Code Rate (CR) match the raw throughput percentage, i.e., by sending an appropriate amount of redundancy a priori with the NC scheme, the raw unreliable throughput is fully utilized while reliability is achieved. Indeed, Fig.~\ref{fig:comparecoderate} shows that the raw throughput percentage closely matches the CRs of the NC-Best cases (dashed line). Also observe that the throughput percentages of HARQ and HARQ-ARQ show large gaps when compared to the best NC schemes, particularly at low PHY code rates. 

At a load of 6~Mbps, HARQ and ARQ do not perform well. Compared to raw, HARQ and HARQ-ARQ show additional losses, reduced throughput and increased delays. Their performances improve as PHY code rates increase, although they do not outperform NC in our experiments. The low performance of HARQ and ARQ may be due to faulty implementation or non-optimal default parameters (e.g., delay timeouts, maximum number of retransmissions). These conclusions, however, require further access to the used equipment to be verified.

Our experimental results suggest that NC has a potential to replace HARQ and ARQ in future wireless network design. We infer that there are three main reasons why NC outperforms HARQ and ARQ. First, NC requires less reliance on ACK packets. In HARQ and ARQ, since the transmitter has to wait for an ACK or NACK packet in each transmission, performance depends on the Round Trip Time (RTT). The longer the RTT, the lower the expected overall performance. In other words, the RTT limits the throughput of HARQ and ARQ. In contrast, in NC, additional degrees of freedom (coded packets) can be sent proactively ahead of time and lost packets can be recovered from successfully received coded packets. Hence, the RTT does not limit the overall performance.

Second, the low throughput of HARQ and HARQ-ARQ suggests that they generate a large amount of overhead. In our experiments, NC overhead reaches 100\% ($N_r=120$) whereas for HARQ, each packet may be retransmitted up to 4 times, as given by HARQ\_MAX\_RETRANSMISSION, thus potentially incurring 400\% overhead. Thus load increases may reach or exceed the supported PHY data rates (see Table~\ref{table:phydatarate}). The excess load may cause buffer overflows at the MAC layer, resulting in a drop in throughput. 

Third, and most importantly, in HARQ and ARQ, each additional redundant packet can only compensate for a particular lost packet, while in NC, each additional coded packet can compensate for any lost packet in the block. In HARQ and ARQ, a particular packet is deemed lost if it is not successfully received after a fixed number of retransmissions. In NC, however, any lost packet can be replaced by the next coded packet. Therefore, NC is more robust to lost packets. It is also less sensitive to lost ACK packets, since it requires at most one ACK packet per block.

\section{Conclusions  \label{sec:conclusions}}

This work proposes and demonstrates a network-coding (NC)--enabled reliability architecture for next generation wireless networks. In our design, NC is used as a packet erasure code providing resilience against errors below the IP layer. We validate our design through an experimental case study at a GENI WiMAX site, where we compare our NC architecture to default HARQ and ARQ in terms of packet loss, throughput and file transfer delay. We demonstrate that NC is potentially superior as a packet erasure code. Compared to HARQ and ARQ, NC potentially offers a gain of 5.9 times in throughput and a reduction of 5.5 times in file transfer delay. Our experimental setups were limited by our ability to access and configure the GENI WiMAX platform. Owing to its flexibility and simplicity, we believe that the proposed NC architecture may become instrumental in providing faster and more efficient next generation wireless network services through low-cost upgrades.

This initial architectural design opens up a number of new and exciting venues for future investigation. Immediate follow-ups may investigate the performance sensitivity to different offered loads. The experimentation could also be extended to investigate various parameters of the proposed design, such as the numbers of redundancy transmission rounds ($N_k$) or concurrent encoder-decoder thread pairs ($N_p$). In addition, wider access to the wireless communication equipment (i.e., WiMAX BS and SS, in our case) would enable a more complete study, encompassing features such as signal-to-noise ratio (SNR) and power control, HARQ and ARQ fine-tuning, operation under adaptive modulation and coding (AMC), and mobility. The optimization of the decoding time is also an interesting direction to pursue. Different decoding algorithms such as the Jacobi iterative method for finite field matrix inversion may be considered. 

Ultimately, the extension of our design to an adaptive scheme that dynamically adjusts various design parameters using information available through ACKs or through channel quality information is an important design goal. In addition, the joint optimization of rate and power control under NC would be a valuable next step. Furthermore, the study may be broadened to mobile SSs, multiple-hop topologies and traffic-dependent coding interfaces.

We believe that the integration of NC within the WMAN protocol stack, namely at the convergence sublayer, will yield additional advantage for both upper and lower layers. For instance, it may alleviate PHY-layer BER constraints or improve responsiveness to channel conditions for various supported traffic flows. Although NC will require upgrades at all participating base and subscriber stations, the involved software and protocol upgrades are minor and well within the reach of operators. Besides, once installed, they may be exploited for service differentiation.

\section*{Acknowledgment}
Our experiments would not have been possible without the technical support of Mr. H.E. Mussman at Raytheon BBN Technologies. The authors would also like to thank Professor Giovanni Pau of UCLA as well as Dr. Danail Traskov, Dr. Ali ParandehGheibi, Dr. MinJi Kim and Jason Cloud for their support and helpful discussions. 

This work is based upon research supported by Orange France Télécom (awards number 0050012310-A100 and 018499-00), the Semiconductor Research Corporation (awards number RA306-S1 and 017894-010), the Air Force Office of Scientific Research (awards number FA9550-09-1-0196, FA9550-08-1-0159, and 016974-002), DARPA (awards number 739532-SLIN 0001 and 739532-SLIN 0002), as well as by FQRNT, Québec's nature and technology research fund. 

 

\newcommand{\noopsort}[1]{} \newcommand{\printfirst}[2]{#1}
  \newcommand{\singleletter}[1]{#1} \newcommand{\switchargs}[2]{#2#1}

\end{document}